# Steric effects of CO$_2$ binding to transition metal-benzene complexes: a first-principles study

Hyeonhu Bae[1], Bing Huang[2] and Hoonkyung Lee[*1]

[1] *Department of Physics, Konkuk University, Seoul 05029, Korea*

[2] *Beijing Computational Science Research Center, Beijing 100194, China*

* *hkiee3@konkuk.ac.kr*

**ABSTRACT**

Using density functional theory (DFT) calculations, we investigated the adsorption of CO$_2$ molecules on 3$d$ transition metal (TM)-benzene complexes. Our calculations show that the maximum number of CO$_2$ molecules adsorbable on Sc or Ti atoms is three, but the 18-electron rule predicts it should be four. The 18-electron rule is generally successful in predicting the maximum H$_2$ adsorption number for TM atoms including Sc or Ti atoms. We found that the 18-electron rule fails to correctly predict CO$_2$ binding on Sc- or Ti-benzene complexes because CO$_2$ binding, in contrast to H$_2$ binding, requires additional consideration for steric hindrance due to the large bond length of CO$_2$. We calculated the occupation function for CO$_2$ using the Tolman cone angle, which shows that three CO$_2$ molecules fully occupy the available space around Sc- and Ti-benzene complexes. This estimation is the same maximum CO$_2$ adsorption number predicted by DFT calculations. Therefore, we propose that the occupation function for CO$_2$ using the Tolman cone angle is an efficient model for evaluating steric hindrance of CO$_2$ adsorption on a surface.

1. **Background**

Since carbon dioxide ($CO_2$) gas causes global warming giving rise to severe climate change [1,2], $CO_2$ capture has been of great interest for mitigating this issue. Recently, nanostructured materials, such as graphene, zeolites, and metal-organic frameworks have received much attention as $CO_2$ capture materials [3-7]. These materials are practically advantageous because of their high $CO_2$ adsorption capacity, fast adsorption kinetics, and effective regeneration. However, these materials exhibit poor selectivity for $CO_2$ in flue gases and low capture capacity at low pressures (~$10^{-3}$ bar) [8-11]. These drawbacks limit the capture of $CO_2$ from flue gases under ambient conditions using conventional capture methods [11]. For reversible $CO_2$ capture under low pressure, the adsorption energy of $CO_2$ molecules should be in the energy window of −1.2 eV to −0.8 eV [12].

More recently, a theoretical study [12,13] showed that transition metal (TM)-porphyrin-like graphene or sheets selectively adsorb $CO_2$ molecules from flue gases with the required adsorption energy. Several $CO_2$ molecules can adsorb on a single TM atom. However, the number and adsorption energy of adsorbed $CO_2$ molecules depend on the adsorption geometry of $CO_2$, which depends on the type of TM atom and the number of $CO_2$ molecules involved. It was found that TM atoms with empty $d$ orbitals can selectively attract $CO_2$ molecules from flue gases. It was theoretically shown that $CO_2$ adsorbs on open TM coordination sites in metal-organic frameworks with enhanced adsorption energy through orbital hybridization between TM atoms and $CO_2$ molecules [14]. Furthermore, Fe–porphyrin-like carbon nanotubes [15] or Co-porphyrin-like nanostructures have been synthesized [16-18]. These materials have shown the possibility that TM-nanostructures can be used for highly selective $CO_2$ capture.

In recent years, TM-$H_2$ complexes [19] have been experimentally prepared in which multiple $H_2$ molecules adsorb on TM atoms accompanied by bond elongation of the $H_2$ molecules. These TM-dihydrogen complexes have been proposed for use as hydrogen storage materials at room temperature and ambient pressure [20-22]. A TM atom can adsorb molecules such as $H_2$ until the TM atom's 4*s*, 4*p*, and 3*d* orbitals are filled with 18 electrons, which is the so-called 18-electron rule [22-24]. Importantly, the 18-electron rule has accurately described the maximum number of attachable $H_2$ molecules on a TM atom, and has been used for predicting the capacity of $H_2$ storage materials [22-24]. In this paper, we investigated adsorption of $CO_2$ molecules on 3*d* TM-benzene complexes using first-principles calculations. Unlike $H_2$ binding on TM atoms, DFT calculations on the maximum number of $CO_2$ molecules attachable to Sc or Ti atoms does not agree with the 18-electron rule, whereas, the rule is obeyed by other TM atoms. This observation is ascribed to the fact that a TM site in the TM-benzene complex can adsorb up to only three $CO_2$ molecules because of steric effects. This steric effect was evaluated using the Tolman cone angle. The 18-electron rule was made to work for Sc and Ti by including a correction term to account for the steric effects related to adding a fourth bound $CO_2$ molecule. This modification of the 18-electron rule makes the number of adsorbed $CO_2$ molecules consistent with the DFT calculation. Our results allow us to predict TM-$CO_2$ complexes for selective $CO_2$ capture materials, based on TM-nanostructures at ambient conditions.

## 2. Computational methods

Our calculations were carried out using the density functional theory (DFT) [25] as implemented in the Vienna ab-initio simulation package (VASP) with the Perdew–Burke–Ernzerhof scheme [26] for the generalized gradient approximation (GGA). The projector augmented wave (PAW) method was used [27]. The kinetic energy cutoff was set to be 800 eV. For calculations of $CO_2$ adsorption, our model for the TM-benzene system comprised a supercell with a vacuum size of 12 Å. Periodic boundary calculations on the model were done.

Geometrical optimization of the TM-benzene system was performed until the Hellmann–Feynman force acting on each atom was less than 0.01 eV/Å.

## 3. Results and discussion

We performed calculations for adsorption of $CO_2$ on TM atoms to investigate the attachment of $CO_2$ molecules on TM-decorated benzene ($C_6H_6$), in which the TM atom was placed in the center of the carbon hexagon. We evaluated Sc, Ti, V, Cr, Mn, Fe, Co, Ni, and Cu atoms as the TM atom. We found that a $CO_2$ molecule adsorbs on a TM atom between the TM atom and the $CO_2$ molecule with a coordination number of one or two as shown in Figure 1(a) and 1(b), respectively. The distance between the $CO_2$ molecule and the TM atom is ~2.0 Å for geometries with one and two coordination numbers. For situations where more than one $CO_2$ molecule is adsorbed on a TM atom, $CO_2$ molecules can have coordination numbers of one or two, or a number between one and two, as shown in Figure 1(c)–1(h).

Conventional hapticity for describing metal-ligand complexes was used to describe the distinct geometries and coordination numbers for $CO_2$ molecules adsorbed on TM atoms. One and two coordination numbers between a TM atom and $CO_2$ molecule are designated as $\eta^1$ and $\eta^2$ configurations, respectively. For example, when a $CO_2$ molecule adsorbs on a Ti-decorated $C_6H_6$ with a coordination number of two, the molecular formula is written as Ti($\eta^6$-$C_6H_6$)($\eta^2$-$CO_2$). Sc($\eta^6$-$C_6H_6$)($\eta^1$-$CO_2$)$_2$($\eta^2$-$CO_2$) is written for the geometry when two $CO_2$ molecules adsorb on Sc-benzene with an $\eta^1$ configuration and a third $CO_2$ molecule adsorbs with an $\eta^2$ configuration. For simplicity, we denote these two examples as 2 and 1+1+2, respectively, as shown in Figure 2(a). Furthermore, there are two types of 2+2 coordinations: one is "para" and the other is "ortho," which indicates symmetric and asymmetric geometries with respect to a plane containing the $CO_2$ molecules, as shown in Figure 2(a), respectively.

The calculated average $CO_2$ adsorption energy (per $CO_2$ molecule) on a TM atom as a function of the number of adsorbed $CO_2$ molecules with a given hapticity is shown in Figure 2(b). Importantly, the $CO_2$ adsorption energy is in the desirable adsorption energy range for reversible $CO_2$ capture at room temperature under low pressure. The calculated adsorption energies and geometries of $CO_2$ molecules are dependent on the type of TM atom. The $CO_2$ adsorption energy for an $\eta^2$ configuration is lower than that for an $\eta^1$ configuration. The 2+2 structures with ortho or para configurations occur when two $CO_2$ molecules adsorb on a TM atom, whose structures are energetically favorable compared with the 1+2 strucures. Moreover, the $CO_2$ adsorption energy is reduced as the number of adsorbed $CO_2$ molecules increases because repulsive interactions between adsorbed $CO_2$ molecules increase (see Figure 2(b)).

We also investigated how to predict the number of $CO_2$ molecules bound to different TM atoms. The number of adsorbed $CO_2$ molecules can be explained by the 18-electron rule [22-24], as was observed for $H_2$ binding to TM atoms. This rule suggests that the TM atom can adsorb several molecules such as $CO_2$ until the TM 4$s$, 4$p$, and 3$d$ orbital shells are fully occupied. Through the analyses involved in our calculations, we found that the number of $CO_2$ molecules per TM atom can be expressed by Eq. (1).

$$_i N_{CO_2}^{18} = [(18 - n_v^i - n_b^i)/2], \qquad (1)$$

where [X] denotes an integer not exceeding X (Gaussian brackets), and $n_v^i$ and $n_b^i$ denote the number of valence electrons of the metal and the number of electrons bonding with the $i$-type TM, respectively. For example, $n_v^{Sc}$ is 3 for Sc atom from the valence electron configuration of $4s^2 3d^1 4p^0$ while $n_b^{Sc}$ is 6 for the six $\pi$ electrons in benzene. From this empirical rule, the maximum number of adsorbed $CO_2$ molecules for Sc, Ti, V, Cr, Mn, Fe, Co, Ni, and Cu is estimated to be 4, 4, 3, 3, 2, 2, 1, 1, and 0 respectively. However, for Sc and

Ti, these numbers (four) do not agree with the results of our DFT calculations (three), while they do agree for the other TM atoms listed.

Next, we investigated the reason why the maximum number of adsorbed $CO_2$ molecules for Sc- and Ti-$CO_2$ complexes, unlike Sc-$H_2$ or Ti-$H_2$ complexes, does not agree with the 18-electron rule. When $CO_2$ molecules bind to a TM adsorption site with limited space, steric effects are not negligible, even though negligible for $H_2$ molecules. This difference is due to the longer bond length (2.20 Å) in a $CO_2$ molecule in a vacuum relative to that of a $H_2$ molecule (0.74 Å). Steric effects result in a reduction of the maximum number of adsorbed $CO_2$ molecules, which is caused by the repulsive interaction between a $CO_2$ molecule and the adsorption site on a TM. The Tolman cone angle [28] was used to estimate the maximum number of adsorbed $CO_2$ molecules resulting from steric effects, as shown in Figure 3. The Tolman cone angle for different $\eta^1$ and $\eta^2$ geometries for $CO_2$ molecules bound to a TM atom is shown in Figures 3(b) and 3(c), respectively. The angle is defined by the tangential lines from the center of the TM atom to the van der Waals radius of the atoms, as shown in Figure 3(b). The van der Waals radii of the O and C atoms [29] were used in this calculation. Using the Tolman cone angle, the occupation function of $CO_2$ for attachable space is defined in Eq. (2).

$$f_i(_i N_{CO_2}) = \frac{\sum_j^{_i N_{CO_2}} \Omega_j^i}{\Omega_0^i}, \qquad (2)$$

where $\Omega_j^i$ denote the Tolman cone (solid) angle between the $j^{th}$ $CO_2$ molecule and the $i$-type TM atom for a given $CO_2$ adsorption number, $_i N_{CO_2}$, respectively. $\Omega_0^i$ indicates the Tolman cone angle of the structure, TM($\eta^6$-$C_6H_6$), without $CO_2$ adsorption, is shown in Figure 3(a). The Tolman cone angle is mathematically given by $\iint \sin\theta d\theta d\phi$, namely,

$\Omega_j^i = 2\pi(1-\cos\theta_j)$. If $f_i \to 1$ $(f_i > 1)$, there is no available space for further $CO_2$ binding.

Next, we calculated the Tolman cone angle for the geometry with the maximum number of absorbed $CO_2$ molecules using the definition given in the previous paragraph. For example, the Tolman cone angle, $\Omega_0^{Sc}$, for $Sc(\eta^6\text{-}C_6H_6)$ is $2\pi \times 1.05$, and $\Omega_1^{Sc}$ is $2\pi \times 0.35$ and $2\pi \times 0.52$ for $Sc(\eta^6\text{-}C_6H_6)(\eta^1\text{-}CO_2)$ and $Sc(\eta^6\text{-}C_6H_6)(\eta^2\text{-}CO_2)$, respectively. The dependence of the angles on the identity of the TM atom was negligible. The calculated value of $f_{Sc}$ for the $\eta^1$ and $\eta^2$ configurations is 0.33 and 0.50, respectively. This means that a $CO_2$ for the $\eta^1$ and $\eta^2$ configurations occupy 33% and 50% of the available space, respectively. For the multiple adsorption case of $Sc(\eta^6\text{-}C_6H_6)(\eta^1\text{-}CO_2)_3$, the total Tolman cone angle for the $CO_2$ molecules was calculated to be $2\pi \times 1.22$. Thus the occupation function, $f_{Sc}$, is 0.83, which means there is no available space for a fourth $CO_2$ because occupation per $CO_2$ molecule requires ~30–50% available space. Thus, because of steric effects in the Sc- and Ti-benzene complexes, the maximum number of adsorbed $CO_2$ molecules is three, smaller by one than the 18-electron rule's prediction. Through calculations of the Tolman cone angle, the occupation functions were determined for all structures evaluated. The maximum number of adsorbed $CO_2$ molecules for each of these structures is listed in Table 1. However, some values of the occupation functions slightly exceed unity. The value of the occupation function can be slightly altered by adjusting the values of the van der Waals radii of the atoms.

We also performed calculations for the adsorption of $H_2$ on Sc. For $Sc(\eta^6\text{-}C_6H_6)(\eta^2\text{-}H_2)_4$ (Figure 4(a)), the occupation function was calculated to be 0.84, indicating that, unlike $CO_2$ adsorption, more than four $H_2$ molecules can be adsorbed. According to a recent study [20,22], up to five $H_2$ molecules can adsorb on Sc bound to a carbon pentagon, which agrees with the 18-electron rule's prediction. From these results, we conclude that the steric effect between $H_2$ molecules does not impact the maximum number of $H_2$ molecules adsorbed.

The 18-electron rule can be modified to account for steric effects by subtracting the correction term, $\gamma_i$, from Eq. (1). The resulting modified 18-electron rule is given by Eq. (3).

$$_i\text{N}_{\text{CO}_2}^{18m} = [(18 - n_v^i - n_b^i)/2] - \gamma_i, \qquad (3)$$

where $\gamma_i$ is the correction value for steric and other effects. The correction term, $\gamma_i$, is calculated from the difference between the maximum adsorption number for $CO_2$ determined from the 18-electron rule and the DFT calculations, namely, $\gamma_i = \text{N}_{\text{CO}_2}^{18} - \text{N}_{\text{CO}_2}^{\text{DFT}}$. The values of the correction term for Sc, Ti, V, Cr, Mn, Fe, Co, Ni, and Cu are estimated to be 1, 1, 0, 0, 0, 0, −1, 0, and 0, respectively. For the Co atom, the number of $CO_2$ molecules adsorbed according to the DFT calculation is still inconsistent with the 18-electron rule's prediction. Two $CO_2$ molecules can be adsorbed according to the DFT calculation and one $CO_2$ molecule according to the 18-electron rule. However, since the adsorption energy (−1.61 eV) of one $CO_2$ molecule is lower than the $CO_2$ adsorption energy (−1.34 eV) for two adsorbed $CO_2$ molecules, the second $CO_2$ adsorption is meta-stable and, hence, negligible. Therefore, the modified 18-electron rule works well for estimating the maximum adsorption number of $CO_2$ molecules for TM-$CO_2$ complexes.

We confirmed that the height change between TM atoms and benzene is negligible regardless of where the TM atom is placed on the carbon pentagon, hexagon, and heptagon. Thus, the occupation function is almost independent of the backbone structure. This calculation indicates that the maximum number of $CO_2$ molecules adsorbed on Sc or Ti is three regardless of whether the backbone structure for carbon is a hexagon or pentagon, which is consistent with a recent study [12] on $CO_2$ binding to TM-porphyrin-like graphene. This prediction is consistent with the results of our DFT calculations as shown in Figures 4(b) and 4(c). However, to describe the maximum number of adsorbed $CO_2$ molecules using Eq. (3), the correction term, $\gamma_i$ for a carbon hexagon and other carbon ring structures will be different based on the number of electrons bonded with the TM atom.

We have discussed the steric effects on the adsorption of gas molecules to TM atoms. The steric effects of $CO_2$ molecules are not negligible, whereas the steric effects of $H_2$ molecules are negligible [22]. This difference is because the bond length of $CO_2$ is much longer than that of the $H_2$ molecule. Thus, for other gas molecules such as $N_2$ and $O_2$, steric effects could influence their adsorption on TM atoms. The 18-electron rule can be corrected with a value of the occupation function, $f_i$, for gas molecules other than $H_2$. The modified 18-electron rule can be used for describing any molecule that consists of atoms close to the van der Waals radius of oxygen. The modified 18-electron rule can be used for estimating the $CO_2$ capacity of any nanostructure containing carbon hexagons decorated with TM atoms where the TM atoms are located on top of the carbon hexagons. In addition, TM-$CO_2$ complexes with $\eta^1$ and $\eta^2$ configurations were confirmed in experiments [30,31], which is consistent with our calculation results. Therefore, the modified 18-electron rule can be employed to predict the adsorption capacity of novel $CO_2$ capture materials, based on TM-decorated nanostructures such as graphene and carbon nanotubes.

4. **Conclusions**

In conclusion, we performed first-principles total energy calculations for $CO_2$ adsorption on TM–benzene. Using an analysis of the occupation function for $CO_2$ to calculate the available $CO_2$ space using the Tolman cone angle, we found that a maximum of three $CO_2$ molecules can adsorb on Sc or Ti atoms because of steric effects. We also proposed an occupation function for quantifying the steric hindrance for gas adsorption on a surface. Our results provide a new approach to understanding steric effects for adsorption of carbon dioxide gas as well as hydrogen gas storage.

**Authors' contributions**

HL conceived and designed the study. BH performed the calculations. BH and HL interpreted the data. All authors revised the manuscript and approved the final version of the manuscript.

**Competing financial interests**

The authors declare no competing financial interests.

**Acknowledgements**

This paper was supported by Konkuk University in 2016.

# Figures

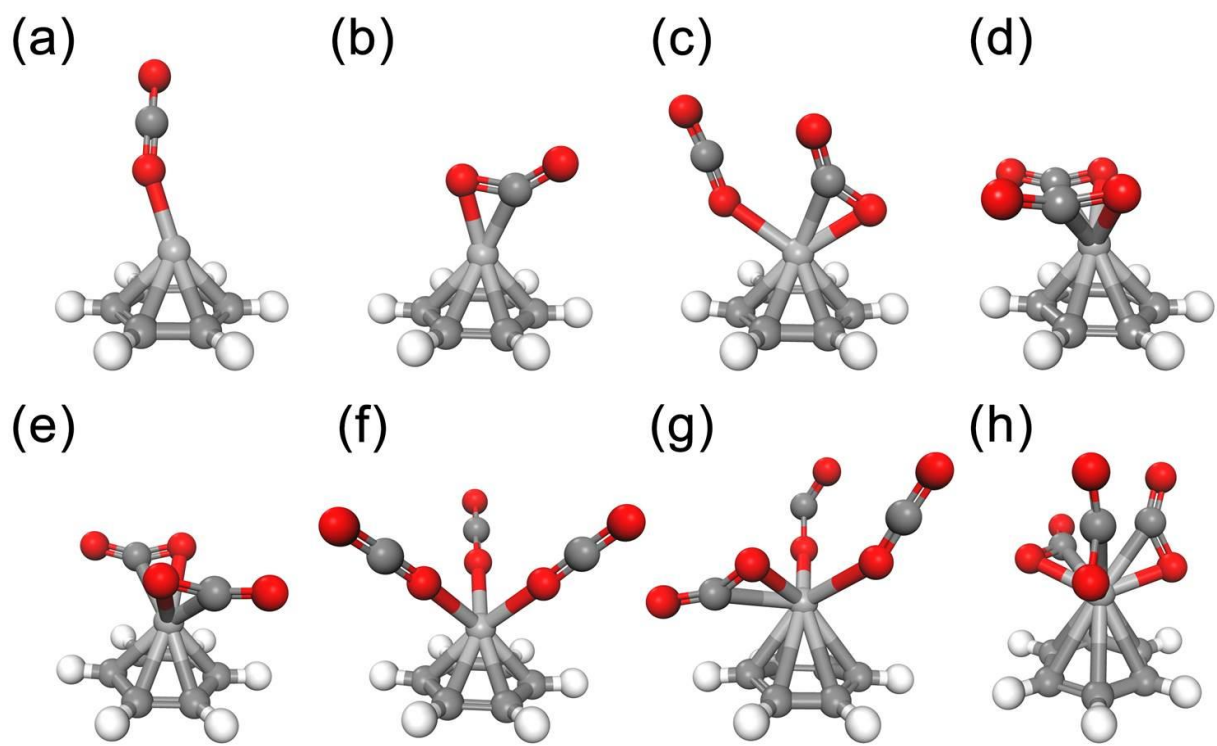

**Figure 1** Structures of TM-benzene complexes with attachment of different numbers and geometries of $CO_2$ molecules. (a) $CO_2$ molecule adsorbed on a TM atom with $\eta^1$ configuration, (b) $\eta^2$ configuration, and (c)−(h) adsorption of multiple $CO_2$ molecules with different configurations.

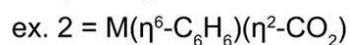
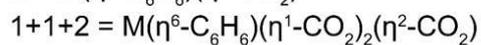
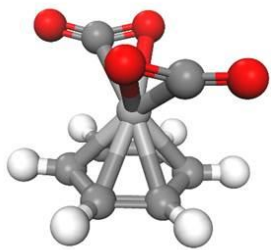
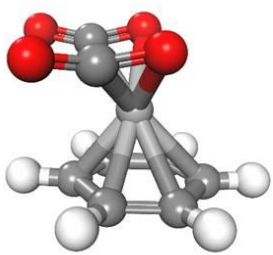
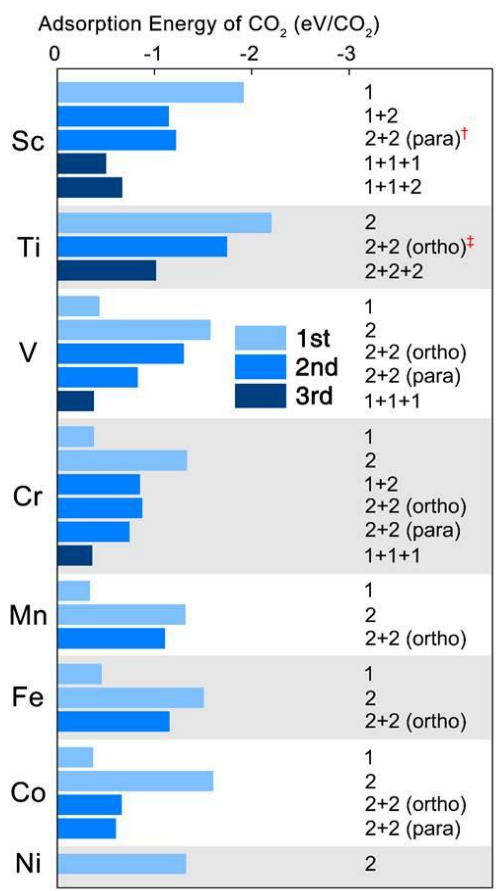

**Figure 2** (a) Top: Examples of hapticity. 1 and 2 indicate the geometric configurations of $\eta^1$ and $\eta^2$, respectively. Bottom: Example of a "para" (symmetric) and "ortho" (asymmetric) configuration. (b) Calculated (average) adsorption energies of $CO_2$ molecules on different TM atoms in TM-benzene complexes. Adsorption of more than one $CO_2$ molecule with mixed configurations of $\eta^1$ and $\eta^2$ are shown in the bars.

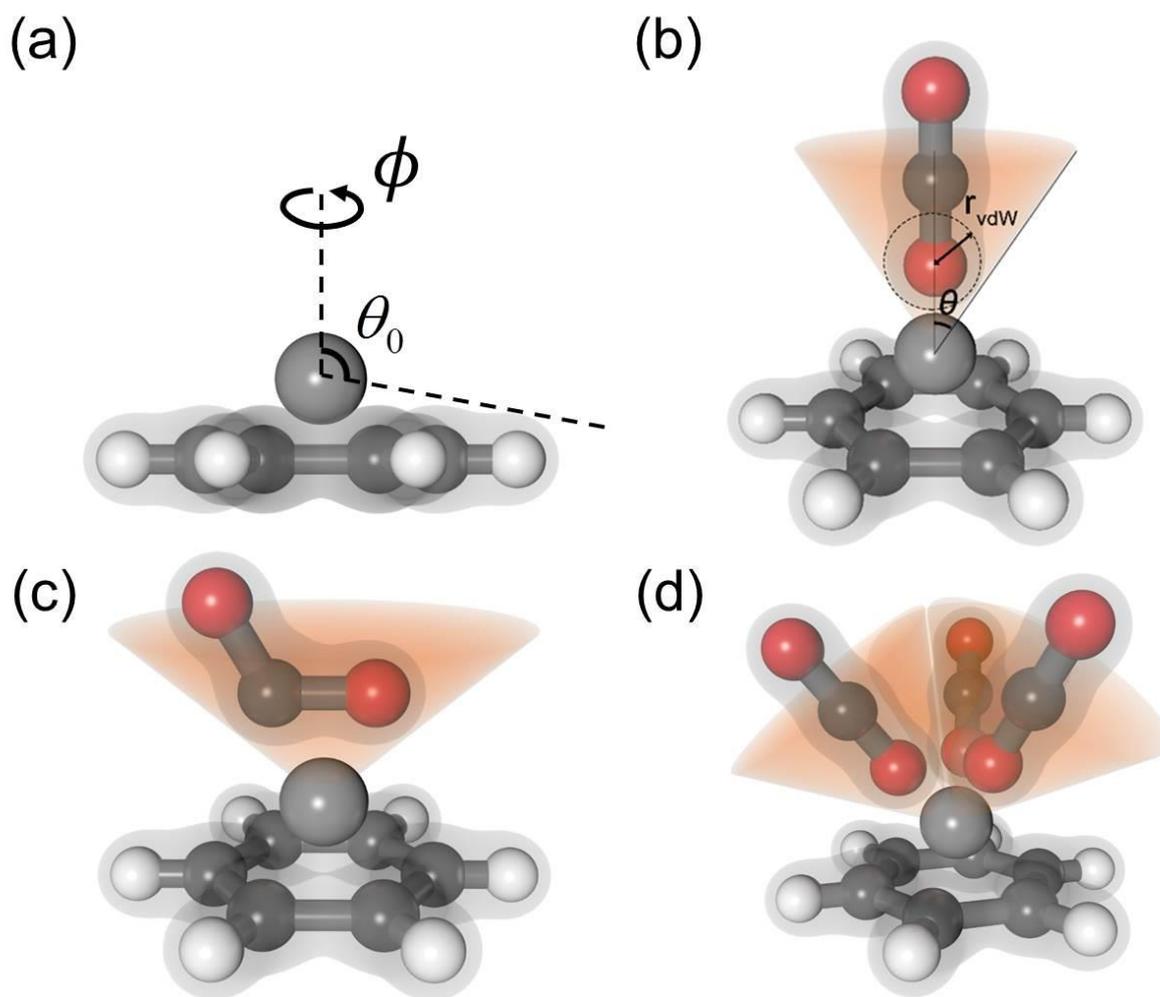

**Figure 3** Definition of the Tolman cone for TM-benzene complexes with attachment of $CO_2$ molecules to TM. (a) Zenith angle, $\theta_0$, on a TM-benzene complex. (b) The dotted curve shows a circle with the van der Waals radius ($r_{vdW}$) of the O atom. $\theta$ denotes the zenith angle for the Tolman cone. (b), (c), and (d) Tolman cone for TM($\eta^6$-$C_6H_6$)($\eta^1$-$CO_2$), TM($\eta^6$-$C_6H_6$)($\eta^2$-$CO_2$), and TM($\eta^6$-$C_6H_6$)($\eta^1$-$CO_2$)$_3$, respectively. Red-shaded surfaces indicate the Tolman cones.

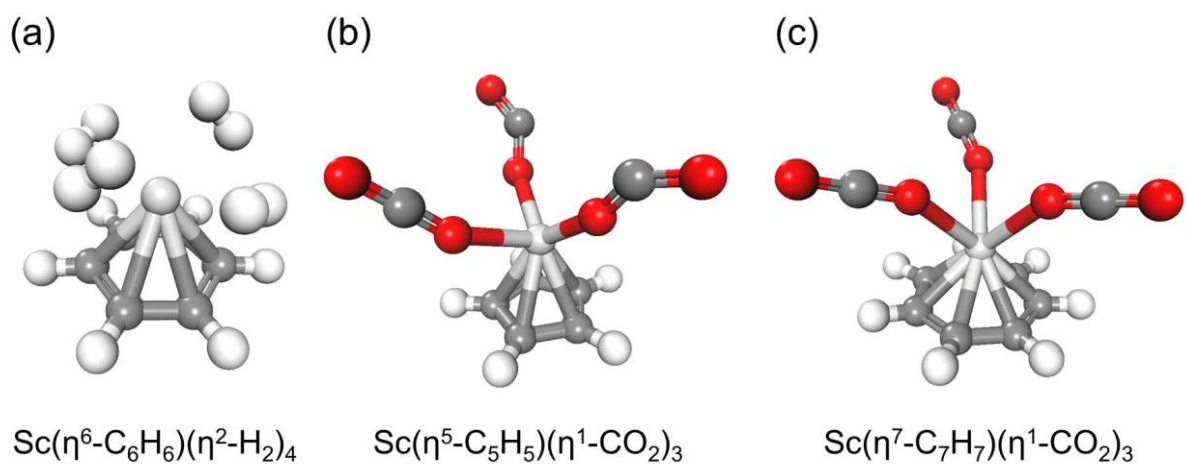

**Figure 4** Structures of Sc-carbon ring complexes with attachment of different numbers and geometries of $H_2$ or $CO_2$ molecules. (a) Four $H_2$ molecules adsorbed on Sc-$C_6H_6$ complex with $\eta^2$ configuration. Three $CO_2$ molecules adsorbed on (b) Sc-$C_5H_5$ complex and (b) Sc-$C_7H_7$ complex with $\eta^1$ configuration.

| TM | η | $\Omega^i_j/2\pi$ | $\Omega^i_0/2\pi$ | $f_i$ | $_i N^{DFT}_{CO_2}$ | $_i N^{18}_{CO_2}$ |
|---|---|---|---|---|---|---|
| Sc | $\eta^1$ | 0.37 | 1.22 | 0.83 | 3 | 4 |
|    | $\eta^1$ | 0.20 | | | | |
|    | $\eta^2$ | 0.44 | | | | |
| Ti | $\eta^2$ | 0.50 | 1.15 | 1.28 | 3 | 4 |
|    | $\eta^2$ | 0.51 | | | | |
|    | $\eta^2$ | 0.47 | | | | |
| V  | $\eta^1$ | 0.27 | 1.11 | 0.74 | 3 | 3 |
|    | $\eta^1$ | 0.27 | | | | |
|    | $\eta^1$ | 0.28 | | | | |
| Cr | $\eta^1$ | 0.30 | 1.10 | 0.81 | 3 | 3 |
|    | $\eta^1$ | 0.29 | | | | |
|    | $\eta^1$ | 0.30 | | | | |
| Mn | $\eta^2$ | 0.57 | 1.10 | 1.04 | 2 | 2 |
|    | $\eta^2$ | 0.57 | | | | |
| Fe | $\eta^2$ | 0.58 | 1.10 | 1.06 | 2 | 2 |
|    | $\eta^2$ | 0.58 | | | | |
| Co | $\eta^2$ | 0.56 | 1.10 | 1.02 | 2 | 1 |
|    | $\eta^2$ | 0.56 | | | | |
| Ni | $\eta^2$ | 0.63 | 1.10 | 0.57 | 1 | 1 |

**Table 1** Calculation of the Tolman cone angle, the occupation functions, and comparison of the number of $CO_2$ molecules adsorbed on TM-benzene complexes by prediction using DFT results and the 18-electron rule. The occupation function for $CO_2$, $f_i$, was calculated from $\sum \Omega^i_j / \Omega^i_0$.